\documentstyle[prl,aps,multicol,epsf]{revtex}

\begin{document}

\draft

\title{Glassy transition in a disordered model for the RNA secondary
structure}

\author{A. Pagnani$^a$, G. Parisi$^b$ and F. Ricci-Tersenghi$^c$}

\address{$^a$ Dipartimento di Fisica and INFM,
Universit\`a di Roma {\it Tor Vergata},
Via della Ricerca Scientifica 1, 00133 Roma (Italy)\\
$^b$ Dipartimento di Fisica and INFN,
Universit\`a di Roma ``La Sapienza'',
Piazzale Aldo Moro 2, 00185 Roma (Italy)\\
$^c$ Abdus Salam ICTP, Condensed Matter Group,
Strada Costiera 11, P.O. Box 586, 34100 Trieste (Italy)}

\date{\today}
\maketitle

\begin{abstract}
We numerically study a disordered model for the RNA secondary
structure and we find that it undergoes a phase transition, with a
breaking of the replica symmetry in the low temperature region (like
in spin glasses).  Our results are based on the exact evaluation of
the partition function.
\end{abstract}


\begin{multicols}{2}
\narrowtext

The folded structure of biopolymers, like RNA and proteins, is crucial
for understanding the biological functionality of these
molecules~\cite{BRANDEN_TOOZE} and its characterization still remains
a challenging problem in statistical mechanics and theoretical
biology~\cite{MECC_STAT}.  The folding problem usually consists in
understanding if and how a particular biomolecule (maybe one selected
by evolution and present now in nature) folds into its native
conformation.  In this Letter we are interested in the
characterization of the most generic (i.e.\ random) RNA molecules.
Even if real RNAs are not completely random, they present a very large
variability in their sequences and no strong correlations in their
bases.  The interest in studying the limiting and somehow unphysical
case of really random sequences arises in order to answer the
following questions.  Is the folding transition, that forces real
biomolecules into their functional shapes, characteristic of those
sequences selected by the evolution?  Do random sequences show some
phase transition too?  We answer affirmatively to both questions,
showing that the transition depends more on the geometrical
constraints and on the interaction energies spread rather than on the
specific sequence.  However in the random case the transition is of a
glassy type and the low-temperature phase is not dominated by a
single native state.  Our results may be very useful in order to
understand better what could happen in a prebiotic world mainly made
of random RNA sequences~\cite{RNA_WORLD}.  Such a transition
(partially found only in a very simplified model of
proteins~\cite{DILL}) was suggested in previous studies of the RNA
folding~\cite{HIGGS}.

In this Letter we first study the thermodynamical properties of random
RNAs, finding some hints for the existence of a glassy transition.
The clear evidence for such a transition is shown in the last part of
the paper and has been obtained thanks to the typical tools of
disorder systems statistical mechanics: spin glass susceptibility and
a related parameter (see Fig.~\ref{fig3}).  The connection with
complex systems is well expected: the model has both disorder and
frustration.

Generally speaking a classification among biopolymers includes a
hierarchy of structures and in principle a complete description must
include all these levels.  RNA from this point of view is supposed to
be simpler than DNA or proteins since its secondary structure seems to
capture the essential features of the thermodynamics of the molecule.
RNA molecules are linear chains consisting of a sequence of four
different bases: adenine ($A$), cytosine ($C$), guanine ($G$) and
uracil ($U$).  The four bases are related by complementarity
relations: $C-G$ and $A-U$ form stable base pairs with the formation
of hydrogen bonds and are also known as Watson-Crick base pairs.

The secondary structure of RNA is the set of base pairs that occur in
its three-dimensional structure. Let us define a sequence as
${\mathcal{R}}\equiv\{r_1,r_2,...,r_n\}$, $r_{i}$ being the $i^{th}$
base and $r_{i}\in\{A,C,G,U\}$. A secondary structure on $\mathcal{R}$
is now defined as a set $\mathcal{S}$ of $(i,j)$ pairs (with the
convention that $1 \leq i \leq j \leq n$) according to the following
rules:

a) $j-i \geq 4\;$: this restriction permits flexibility of the chain
in its three-dimensional arrangement.

b) Two different base pairs $(i,j),(i',j') \in \mathcal{S}$ if and
only if (assuming with no loss of generality that $i<i'$):

$i<j<i'<j'\;$: the pair $(i,j)$ precedes $(i',j')$;

$i<i'<j'<j\;$: the pair $(i,j)$ includes $(i',j')$.

Condition b) avoids the formation of pseudo-knots on the structure and
the resulting structure can be drawn on a plane. In real RNA
structures it is known that pseudo-knots occur but are rare and they
can be excluded as a first approximation~\cite{NUSS_JACOB}.

The energy of a structure is simply defined as $H[{\mathcal{S}}] =
\sum_{(i,j)\in{\mathcal{S}}} e(r_{i},r_{j})$.  Other phenomenological
parameters (including stacking energies and loop penalties) could be
considered in order to take into account the whole complexity of the
energy function~\cite{VIENNA}.

In our approach we assume a drastic approximation to the original
model in order to improve its tractability both from numerical and
analytical point of view. As a first step we consider sequences of
only two symbols $(A,B)$, that appear with equal probabilities, and we
assume that only two kind of base pairs occurs: $A-A$ and $B-B$ pairs
with energy $-1$ (in arbitrary units); $A-B$ and $B-A$ pairs with
energy $-2$.  It is reasonable to assume that such a reduction of
symbols will not affect the thermodynamical class of criticality of
the model (this claim is supported by numerical results we have
obtained with a 4-letter code and Watson-Crick base pairs).  We did
not remove the constraint which forbids the links on short distances,
but we simplify it to: $j-i \geq 2$.  We think that this topological
constraint must be kept in order to not drastically change the entropy
of the model and then its thermodynamical behavior.  In this model
disorder (encoded in the sequence ${\cal R}$) and frustration (induced
by the planarity condition on ${\cal S}$) are clearly distinct.  We
hope this could make the model analytically more manageable.

The model can be formally considered as unidimensional and with long
range interactions: the disorder giving rise to different interactions
strengths, all with the same sign (here the disorder does not induce
frustration), while the planarity condition making the long-distance
links unlikely.  We have numerically estimated that the probability of
having a link between two bases a distance $r$ apart goes down roughly
like $r^{-3/2}$.

The planar structure of the configurations and the simple energy
function chosen allow to write down~\cite{NUSS_JACOB} a recursion
equation for the partition function of the subsequence contained
inside the base interval $(i,j)$:
\begin{equation}
Z_{i,j} = Z_{i+1,j}+ \sum_{k=i+1}^j Z_{i+1,k-1} e^{-\beta e(i,k)}
Z_{k+1,j} \quad ,
\label{ZETA}
\end{equation}
with $Z_{i,i}=Z_{i,i-1}=1 \; \forall i$.  Such a recursion relation is
particularly effective since the time needed for the computation of
$Z_{1,L}$ scales as $O(L^3)$.  With a slight modification of the
algorithm it is also possible to include similar recursions for the
internal energy $U = \langle H[\mathcal{S}] \rangle$ and its second
moment $U^{(2)} = \langle H^2[\mathcal{S}] \rangle$, where $\langle
\cdot \rangle$ is the usual average over the Gibbs-Boltzmann
distribution.  At this level all the observables actually depend on
the sequence over which they have been calculated and, if we want to
gain information on the class of universality of the model, we have to
average them over all the random realizations of the sequence.

\begin{figure}
\epsfxsize=0.9\columnwidth
\hspace{0.01\columnwidth}\epsffile{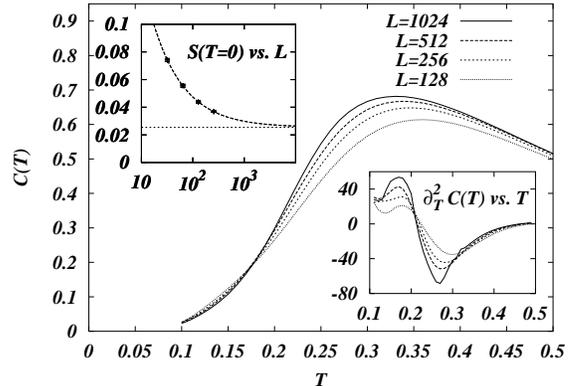}
\caption{The specific heat (and its second derivative in the lower
inset) as a function of the temperature for different sizes.  Upper
inset: zero-temperature entropy versus size and the best power law
fit.}
\label{fig1}
\end{figure}

In Fig.~\ref{fig1} we show the specific heat (averaged over the
disorder) for sizes ranging from $L\!=\!128$ to $L\!=\!1024$.  We note
a very slow increasing of the peak height with the size, which seems
not to diverge.  There is no hint for a finite jump in $C(T)$.  This
could be compared with the result by Bundschuh and Hwa~\cite{HWA} who
found a finite jump in the specific heat (note however that their
model has an unique ground state, which dominates the frozen phase).
It is important to point out that in the temperature region $T \simeq
0.15 - 0.2$ the curves slightly cross themselves and as a consequence
the decrease of $C(T)$ becomes steeper for larger sizes.  One of the
main effects of the disorder is that the location on the temperature
axis of the critical region becomes sample-dependent.  A measure of
the critical region width can be achieved from the sample to sample
fluctuation of the temperature where the specific heat has a peak
($\Delta T_p$).  We find that $\Delta T_p \propto L^{-\omega}$, with
$\omega = 0.26$.  If we assume that these fluctuations are induced by
the presence of a nearby transition, we obtain a value
$\nu=\omega^{-1} = 3.9(1)$.

Since the model is unidimensional,
$\alpha\!=\!2\!-\!d\,\nu\!=\!-1.9(1)$ and then the second derivative
of the specific heat with respect to the temperature should display a
very slow divergence or a finite jump.  In fact, in the lower inset of
Fig.~\ref{fig1}, can be seen that the argument is fully supported by
the data, which show the typical finite size behavior of a
discontinuity.  The clear crossing point of the data around $T \simeq
0.2$ is supposed to be a signature for non-analyticities in
thermodynamical potential.  We note that such a point is located well
below the peak temperature.  This is a common feature in many
disordered systems (e.g. spin glasses).  Near this temperature also
the entropy of the model has a crossing point, which signals a rapid
shrink of the available phase space.

Moreover the model has a finite zero-temperature entropy (see upper
inset in Fig.~\ref{fig1}).  The zero-temperature results have been
obtain via an exact enumeration of all the ground states structures
(GSS) for any given sequence.  The number of GSS (i.e.\ the
degeneracy) strongly depends on the sequence: for example, studying
thousands of different sequences with length $L=256$, we have found
sequences with degeneracies ranging from 1 to ${\cal O}(10^7)$.  In
the upper inset of Fig.~\ref{fig1} we show the zero-temperature
entropy defined as $S(T=0) = \log({\cal N})/L$, where ${\cal N}$ is
the GSS degeneracy and $L$ is the sequence length, as a function of
$L$.  The line is the power law extrapolation, which tends to
$S(T=0)=0.0255(8)$~\cite{NOTE1}.

Since the model turns out to be highly degenerate in the
low-temperature phase, the natural question is how these GSS are
organized.  It is quite obvious that a very different physical
behavior may appear in a model whose GSS are all very similar (like an
ordered or ``ferromagnetic'' behavior) compared to a model whose GSS
are sparse over the whole configurational space.  A more quantitative
analysis can be achieved introducing the notion of distance between
structures and a classification based on these distances.  To quantify
the relative distance between two structures, we have used the
overlap, which is defined as
\begin{equation}
q[{\cal S},{\cal S}'] = \frac1L \sum_{i<j} l_{ij}^{({\cal S})}
l_{ij}^{({\cal S}')} \quad ,
\end{equation}
where the variable $l_{ij}^{({\cal S})}$ ($l_{ij}^{({\cal S}')}$)
takes value 1 if sites $i$ and $j$ are connected in the ${\cal S}$
(${\cal S}'$) sequence and 0 otherwise.  By definition the overlap
takes values in the interval $[0,1]$.  For any given disorder
realization (i.e.\ sequence) ${\cal R}$ we can define the
zero-temperature probability distribution function (pdf) of the
overlaps as
\begin{equation}
P_{\cal R}(q)=\sum_{{\cal S},{\cal S}'\in \Gamma_{\cal R}}
\delta(q-q[{\cal S},{\cal S}']) \quad ,
\end{equation}
being $\Gamma_{\cal R}$ the GSS set.  This definition can be easily
generalized to every temperature summing over all the structures and
weighting each term with the Gibbs-Boltzmann factor of ${\cal S}$ and
${\cal S}'$.  The usual classification of disordered
systems~\cite{MPV} is based upon the average pdf of the overlaps, the
so-called $P(q)\equiv [P_{\cal R}(q)]$, the average being taken over
the disorder distribution function.

We have calculated the $P(q)$ at different temperatures,
$T\in[0,0.4]$.  While at $T=0$ we summed over the whole sets
$\Gamma_{\cal R}$, at finite temperatures we performed a Monte Carlo
sampling of the structures in the spirit of Higgs~\cite{HIGGS}.

\begin{figure}
\epsfxsize=0.9\columnwidth
\hspace{0.01\columnwidth}\epsffile{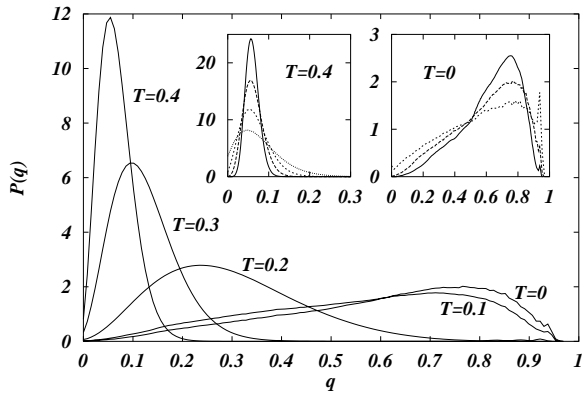}
\caption{The $P(q)$ for different temperatures. Insets: the size
dependence of $P(q)$ in the high (left) and low (right) temperature
phases.}
\label{fig2}
\end{figure}

In Fig.~\ref{fig2} the averaged $P(q)$ are shown.  The first striking
evidence is that, decreasing the temperature, the shape of the $P(q)$
changes abruptly from a narrow peak in the low-$q$ region to a broader
one which extends over almost the whole allowed support.  In the
insets we present the size dependence of $P(q)$ for the highest and
lowest temperature considered.  For the $T=0.4$ case, we are highly
confident that the thermodynamical limit would be a delta function
(the width of the distribution goes to zero as $\Delta q \propto
L^{-1/2}$).  For the $T=0$ case, the asymptotic shape is much more
difficult to be extrapolated, since the width of the $P(q)$ scales
with a small power of $L$ (as in~\cite{HIGGS}) and, eventually, we can
not exclude that it goes to a finite value, implying a breaking of the
replica symmetry.

While in Fig.~\ref{fig2} the averaged $P(q)$ gives us information
about the typical pdf of the overlaps, we can get some hints about the
origin of the $P(q)$ broadness in the low-temperature phase analyzing
directly the $P_{\cal R}(q)$ for each sequence.  If all the GSS of a
given sample are very similar its $P_{\cal R}(q)$ will be non-zero
only in a narrow $q$-range not too far from the upper bound $q=1$.  On
the other hand, if the GSS are very heterogeneous their mutual
overlaps will cover a large $q$-range.

The great majority of the sequences shows a very broad $P_{\cal
R}(q)$, signaling a strong heterogeneity in the GSS.  Moreover the
shape of the pdf completely changes from sequence to sequence (this
property is called non-self-averageness in spin glass
jargon~\cite{MPV}).  Nevertheless some patterns can be easily
recognized: while single peak shapes are mostly associated with
low-degeneracy sequences, highly structured ones seem to be not
correlated to their degeneracy and they are responsible for the $P(q)$
broadness.  Among the latter the double-peak shape dominates,
especially for the sequences with higher entropy: the higher $q$ peak
gives information about the typical distance between two structures in
the same state~\cite{NOTE2}, while the lower $q$ one can be associated
with the rising of a backbone~\cite{ZECCHINA}, that is the set of
persistent links common to all the GSS (already found
in~\cite{MORGAN}).  The position of this second peak strongly
fluctuates from sample to sample, giving rise to the long tail in the
$P(q)$, like spin glass models in external field.

In order to understand whether a true transition happens in this
model, we have measured the order parameter introduced
in~\cite{MANAPAPIRIZU}
\begin{equation}
A = \frac{[\chi_{\cal R}^2]-[\chi_{\cal R}]^2}{[\chi_{\cal R}]^2}
\quad ,
\end{equation}
where $\chi_{\cal R} = L (\Delta q_{\cal R})^2$, being $\Delta q_{\cal
R}$ the width of $P_{\cal R}(q)$.  The $A$ parameter measures how much
the $P_{\cal R}(q)$ changes from sample to sample.  The crossing point
of different curves in Fig.~\ref{fig3} signals the existence of a
low-temperature spin glass phase, where the $P_{\cal R}(q)$ become
non-self-averaging (analogous results has been obtained with the
4-letter model).  In this phase the RNA is ``folded'', that is the
number of links is nearly the maximum allowed.

The critical temperature of this transition seems to be located
between $T=0.1$ and $T=0.15$.  We have determined the best estimates
for $T_c$ and for the critical exponent $\eta$ requiring the best
collapse for the susceptibility $\chi \equiv [\chi_{\cal R}]$ data,
scaled assuming the usual finite-size formula $\chi = L^{2-\eta}
f(L^{-1/\nu}(T-T_c))$ (see inset of Fig.~\ref{fig3}).
We obtain the values $T_c\simeq 0.13$ and $\eta \simeq 1.41$.  We
stress that we also tried to collapse the data fixing $T_c=0$, but the
result was very poor.

\begin{figure}
\epsfxsize=0.9\columnwidth
\hspace{0.01\columnwidth}\epsffile{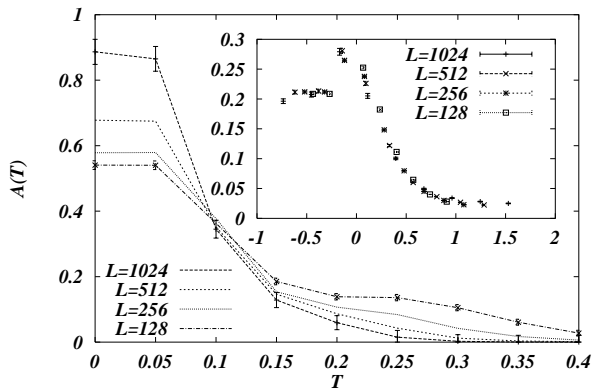}
\caption{The crossing of the $A$ parameter signals the transition to a
glassy phase with the replica symmetry broken. Inset: scaling plot of
$\chi L^{\eta-2}$ versus $L^{-1/\nu}(T-T_c)$.}
\label{fig3}
\end{figure}

The critical temperature seems to be below the one we found by the
study of thermodynamical quantities. However given the high value of
$\nu$, the critical region should shrink as $L^{-1/\nu}$ and then all
the region around $T=0.1-0.3$ is critical as suggested by the wide
separation of the two peaks in $\partial^{2}_{T}C$ (lower inset of
Fig.~\ref{fig1}).

We have presented strong numerical evidences for a phase transition in
a random model for the RNA secondary structure.  It is very important
to stress that the thermodynamical limit is not so interesting for
biological RNAs, which are at most thousands bases long.  As a
consequence our sizes are in principle directly comparable with a
large number of biological molecules.  Our findings about the broadness
of the $P(q)$ could suggest for the existence of zero-energy
fluctuations of the order of the volume, which is a well known
behavior in spin glass and disordered systems.  In~\cite{MARTIN}, for
example, it has been found that the matching problem (which is
disordered and frustrated) has low-energy excitations of order
$\sqrt{L}$.  These excitations becomes irrelevant in the
thermodynamical limit, but they are a key ingredient in order to
correctly describe finite systems.  In the low temperature region of
our model (from $T=0.13$ down to $T=0$) $\chi \propto L^{0.6}$ and it
has strong fluctuations from sample to sample. This situation can be
described by an effective breaking of the replica symmetry with a
strength which goes to zero as $L^{-0.4}$ according to the slow
shrinking of $P(q)$ at $T=0$.  Incidentally we note that in all this
temperature region the critical exponent of $\chi$ is the same, as
suggested from the scaling plot in the inset of Fig.~\ref{fig3}.
Moreover we have also measured the $G$ cumulants defined
in~\cite{MANAPAPIRIZU} and we have verified that it goes to the value
$\frac13$ as the temperature goes to zero coherently with a replica
symmetry breaking scenario.

In conclusions, we have found a glassy transition in a simplified
random model for the RNA secondary structures.  This transition
corresponds to the breaking of the configurational space in many
disconnected regions (ergodicity breaking).  In terms of random RNA
folding, this means that below the critical temperature almost every
sequence folds (all the low-energy structures are very compact), but
very often not in a single structure.  The ergodicity breaking is of
primary importance also for the folding dynamics, that may become very
slow (glassy).

We have checked that the transition disappears as soon as we remove
the constraint of not having links on short distances (maybe this is a
pathology of the 2-letters code) or as soon as we set all the
interaction energies to the same value.  These facts suggest us that
the glassy transition is mainly due to the freezing of some strong
links, which then force the rest of the interactions, aided by the
geometrical constraints.  A cooperation phenomenon between interaction
energies heterogeneity and geometrical constraints has been already
observed in DNA models~\cite{CULE}.

We warmly thank R. Zecchina for many interesting discussions and for a
careful reading of the manuscript.

\end{multicols}
\end{document}